# A Review on Hydrogen Production Technologies and Its Future Demand

Mohammad Abubakr[1], Suhaib Shahid[2], and Iram Arman[3]

1 Department of Chemical Engineering, Z.H.College of Engg.& Technology, A.M.U., Aligarh, U.P
2 Department of Mechanical Engineering, Z.H.College of Engg & Technology, A .M.U., Aligarh, U.P
3 Department of Chemical Engineering, Z.H.College of Engg.& Technology, A.M.U., Aligarh, U.P
Presenting Author's Email Id: mohammadabubakr037@gmail.com, *mohdsuhaib262@gmail.com*

**Abstract**

Hydrogen production is a vital process in the quest for decarbonization and a sustainable future. This conversation explores the various technologies used in hydrogen production, such as steam methane reforming, electrolysis, and biomass gasification. These methods have different applications and efficiencies, but all contribute to the production of hydrogen gas. The future of hydrogen production looks promising, as it offers a clean and versatile energy source that can be used in various sectors, including transportation, industry, and power generation. The increasing demand for hydrogen in these sectors, coupled with the global push for decarbonization, highlights the importance of advancing hydrogen production technologies and infrastructure.

This paper focuses broadly on different methods of hydrogen production like steam methane reforming, electrolysis, and biomass gasification. Steam methane reforming is currently the most common method, accounting for about 95% of global hydrogen production. Electrolysis is another method that uses electricity to split water into hydrogen and oxygen. Biomass gasification involves converting organic materials into hydrogen gas through a thermochemical process. The renewable energy sources considered are water and biomass and the methods considered are Electrolysis (grid), thermolysis, thermochemical water splitting, photoelectrochemical water splitting, and gasification. Electrolysis of water to produce hydrogen accounts for about 5% of the total hydrogen production.

Statistically, global hydrogen production reached about USD 155.35 billion in 2022, with the majority of it being used in the petroleum refining and ammonia production industries. As for applications, hydrogen can be used as a fuel for fuel cell vehicles, as a feedstock for chemical processes, and as a storage medium for renewable energy. It can also be blended with natural gas for heating and cooking purposes. The potential applications of hydrogen are vast and varied, and its future demand is expected to increase as countries strive to reduce carbon emissions and transition to a more sustainable energy system.

**Keywords:** Electrolysis, Gasification, Hydrogen Production, Renewable energy, Storage

# 1. Introduction

At the moment, energy consumption and supply patterns are clearly not sustainable from an environmental, economic, or social perspective. In the absence of a clear solution, increasing the rate at which fossil fuels are used would increase worries about the safety of energy reserves and human-caused greenhouse gas emissions. In order for governments and commercial sectors to proceed in the correct direction, environmentally friendly energy generation technologies must be implemented according to a clear plan. Therefore, for long-term energy sustainability and global security, decarbonizing the energy supply through alternative clean, sustainable, and renewable energy is crucial [1]. According to the literature, hydrogen-based energy storage systems, also known as the hydrogen economy, are paving the way for a society that uses only renewable energy sources [1].

Hydrogen is considered the finest alternative energy carrier of the future, because of its higher energy density on a mass basis, lower environmental obstacles, existence in various forms throughout the universe, and potential to be transformed into useful chemicals or electricity. Hydrogen is a fuel that may be produced from a variety of sources and used in a multitude of applications throughout the energy industry.

This review paper gives an overview of several different hydrogen production techniques, comprising biomass gasification, electrolysis, and fossil fuel reforming. Electrolysis is considered one of the well-known approaches for hydrogen generation and is regarded as one of the effective methods for water dissociation [2]. Thermochemical water decomposition or thermolysis is a high-temperature process of decomposing water. In Photo Electrochemical hydrogen is produced using one or two photo-electrodes fabricated by semiconductor materials provided to generate electricity through solar energy. Biomass mainly contains Steam gasification and Supercritical water gasification processes for the production of hydrogen. A hydrogen economy has long been promoted as a groundbreaking aspect of a low-carbon future. we project that the global hydrogen demand could reach over 2.3 Gt annually, compared to the 90 Mt per year used today [3].

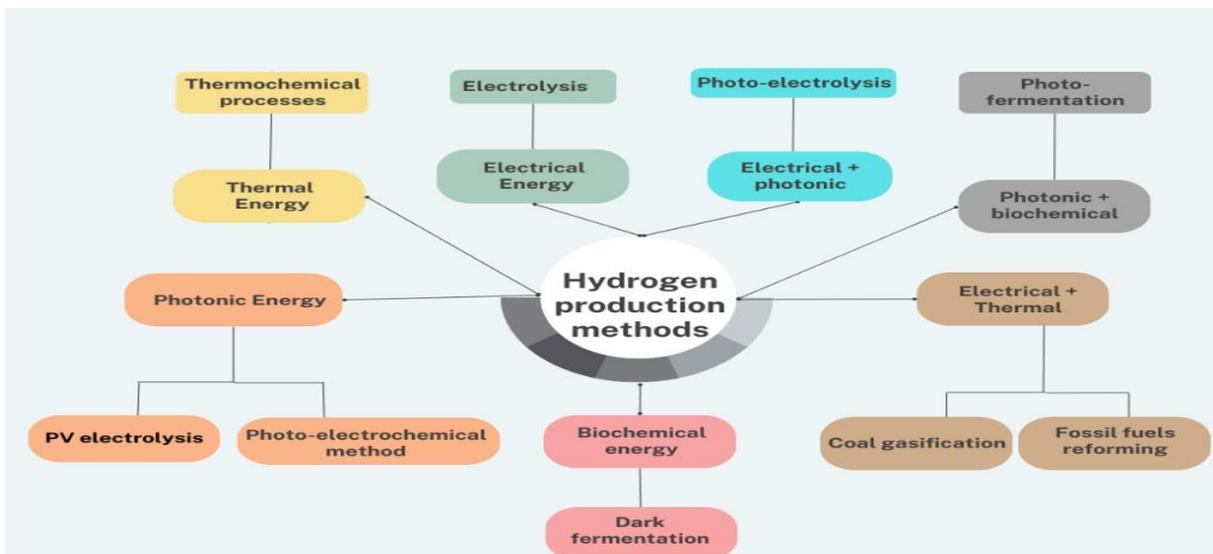

**Fig. 1.** The Hydrogen Production Methods [2].

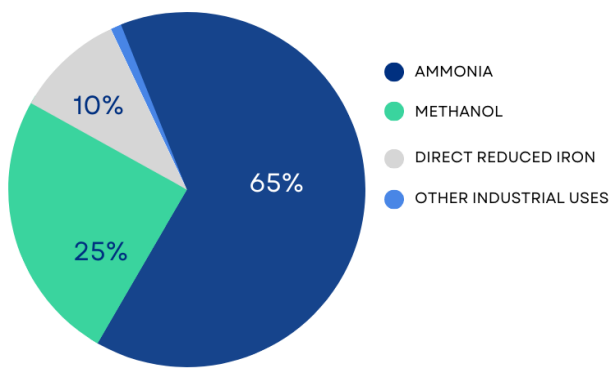

**Fig. 2.** Hydrogen demand in Industry [3].

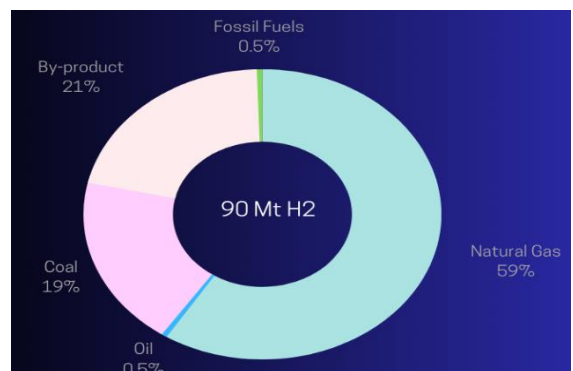

**Fig. 3.** Sources of Hydrogen Production [3].

## 2. Methodologies of Hydrogen Production

### 2.1 Water Electrolysis

Water electrolysis methods are categorized by the electrolyte used to separate the anode (oxygen evolution reaction) and cathode (hydrogen evolution reaction). The main technologies are Alkaline Electrolysis, Polymer Electrolyte Membrane Electrolysis (PEM), and Solid Oxide Electrolysis (SOE).

Alkaline Electrolysis facilitates the breakdown of water at the cathode into hydrogen and HO-. The HO- ions then traverse through the electrolyte and a partitioning diaphragm or membrane, eventually reaching the anode, where they release $O_2$. The electrolyte contains 20-40% NaOH or KOH in water. Operating conditions are 343-363 K at pressures up to 3 MPa [4]. Alkaline electrolysis technology offers several advantages, including its straightforward and user-friendly operation, the absence of the need for costly catalysts, and its extended lifespan. Reported energy efficiency values for alkaline electrolysis have reached as high as 85%, although these figures may vary across different studies, ranging from around 80% to 77%.

The partial reaction at the electrodes is given by:

$2H_2O + 2e^- \longrightarrow H_2 + 2OH^-$    Cathode

$2OH^- \longrightarrow 0.5O_2 + H_2O + 2e^-$    Anode

Proton Exchange Membrane (PEM) electrolysis operates in a corrosive acidic environment, requiring noble metal catalysts like iridium and platinum for anode and cathode reactions. Water is oxidized at the anode, producing $O_2$ and releasing H+ ions, which travel through the membrane and are reduced to hydrogen at the cathode [4]. The polymer electrolyte membrane exhibits remarkably low cross-permeation, resulting in the production of highly pure hydrogen, typically exceeding 99.99% $H_2$ after the hydrogen drying process, which surpasses the purity achieved by AEL [4]. PEM electrolysis has advantages such as safety, environmental cleanliness, and efficiency compared to alkaline electrolysis. However, it's costly with a short lifespan, primarily due to acidic proton exchange membranes and the need for precious metal catalysts.

The following partial reactions take place in this electrolysis:

$2H^+ + 2e^- \longrightarrow 2H_2$    Cathode

$H_2O \longrightarrow 0.5O_2 + 2H^+ + 2e^-$    Anode

High-temperature solid oxide electrolysis (SOE) works at 800-950°C can use heat energy to reduce power consumption [5]. Incoming feed water or steam is preheated in a recuperator, exchanging heat with hot product streams. Steam and recycled hydrogen at the cathode increase reactant utilization (40-80%). The steam-hydrogen mixture is separated through cooling and condensation. To reach the desired SOE inlet temperature (700-1000°C), steam is further superheated, either externally or with an electrical heater [5]. High-temperature solid oxide electrolysis can reduce costs, with potential electricity replacement by external heat sources, while also enhancing electrode activity and overall performance [6].

The reactions at the electrodes are:

$H_2O + 2e^- \longrightarrow H_2 + O^{2-}$    Cathode

$O^{2-} \longrightarrow 0.5O_2 + 2e^-$    Anode

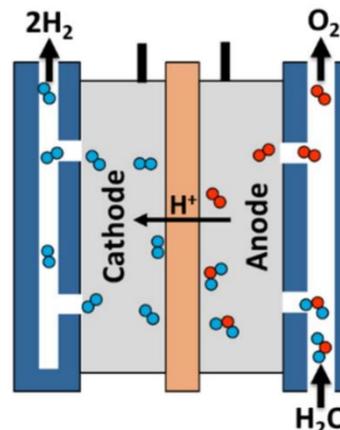

**Fig. 4.** Electrolysis of water.

## 2.2 Thermochemical Water Splitting

Thermochemical water-splitting cycles (TWSCs) revolve around the decomposition of water through a cyclic sequence of chemical reactions that utilize recycled materials and intermediate reactions to achieve the overall reaction that is equal to the dissociation of the water molecule into hydrogen and oxygen [7]. The thermochemical cycles are propelled by thermal energy only, referred to as pure thermochemical cycles, or by thermal energy in conjunction with another energy source (such as electrical or photonic energy), referred to as hybrid thermochemical cycles. Hydrogen and oxygen are the outputs of thermochemical water-splitting cycles [7]. Water may be split down into $H_2$ and $O_2$ in a single step. However, thermochemical cycles are proposed as a repeating set of reactions in which water is divided using thermal energy at temperatures below 2000°C and typically in two or multi-steps [7]. This is because of the poor thermodynamics and the very high temperature necessary for the single-step reaction.

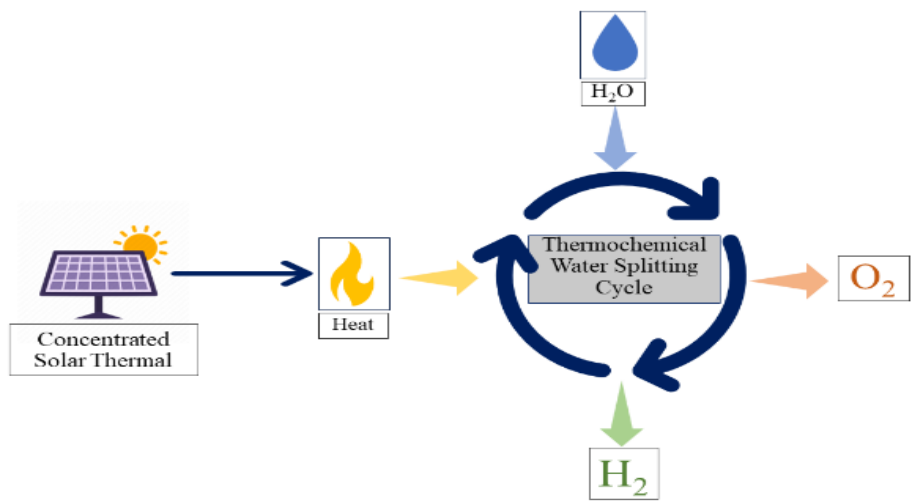

**Fig. 5.** Thermochemical Water Splitting Cycle.

In a two-step thermochemical cycle, a metal oxide is first reduced in an endothermic reaction facilitated by solar radiation, liberating $O_2$. Next, it integrates with water in an exothermic step, forming $H_2$ and the pristine oxide, which is recycled back to the first step [7]. The thermodynamic superiority of TWSCs based on ZnO/Zn redox pairs over other metal oxide redox pairs is well-established, and their processes are realistically tested in solar reactors and thermodynamically analyzed [7].

In a three-step thermochemical cycle, two-step processes can be combined to create three-step processes by substituting a two-step reaction process for the reaction at the greatest temperature. This modification will result in a lower maximum temperature requirement for the cycle. The most well-known three-step cycle is the sulfur-iodine (S-I) cycle, frequently referred to as the general electric cycle. The set of reactions that recur in a three-step thermochemical cycle is summarized as follows [7]:

$H_2O + A \rightarrow 0.5\ O_2$ (or $0.5\ H_2$) + $AH_2$(or AO)

$AH_2$ (or AO) + B $\rightarrow 0.5\ H_2$(or $0.5\ O_2$) + AB

AB $\rightarrow$ A + B

A four-step thermochemical cycle comprised of a hydrolysis procedure, a hydrogen-evolving reaction, an oxygen-evolving reaction, and a reagent-recycling reaction. Research on thermochemical cycles demonstrates that a cycle's maximum temperature requirement for its heat source can be lowered by increasing the number of phases in the cycle. The following is the general form for four-step thermochemical cycle reactions [7]:

$H_2O + AB \rightarrow AH_2 + BO$

$AH_2 \rightarrow H_2 + A$

$BO \rightarrow 0.5 O_2 + B$

$A + B \rightarrow AB$

In hybrid thermochemical water splitting cycles, one reaction is powered by electricity and the other by heat. The electrochemical reaction that is chosen for this step has a lower work consumption than the water electrolysis step. One benefit of hybrid processes is that they can be powered by nuclear waste heat or moderate-temperature heat supplies, whereas two-step processes need significantly lower temperatures [7]. Hybrid methods can achieve up to 48–50% energy efficiency. The most well-known hybrid method for producing thermochemical hydrogen is the two-step hybrid sulfur (HyS) cycle, sometimes referred to as the Westinghouse cycle. The cycle consists of one step of sulfuric acid thermal decomposition, where heat develops and oxygen is consumed, and one step of $SO_2$ electrolysis, where hydrogen is produced and electricity is consumed [8]. Sulfur dioxide electrolysis requires approximately 0.17 V of voltage, which is far less than the 1.23 V of voltage necessary for water electrolysis [8].

## 2.3 Photo-electrochemical method

PEC, or Photoelectrochemical, is a method for producing hydrogen from solar energy by utilizing electrochemical reactions. This technique involves the use of one or two photo electrodes made from semiconductor materials to convert solar energy into electricity. The PEC method has a wide range of applications, including generating electrical current, producing hydrogen, and treating toxic solutions as well [9]. The operation of PECs closely resembles that of photovoltaic (PV) cells, where high-energy photons (exceeding the semiconductor material's band gap) generate electron-hole pairs. These resulting electron-hole pairs are then utilized to catalyze the oxidation and reduction of water. PEC devices conduct both solar energy absorption and water electrolysis simultaneously. Unlike PV systems, PECs don't require a separate power generator system, making them more compact. Thus far, various semiconductor materials have been under investigation and testing for use in PEC systems. Materials like $TiO_2$, $Cu_2O$, $ZnO$, $Fe_2O_3$, $BiVO_4$, and $WO_3$, along with metal oxy-nitrides (such as TaON), n-and p-type silicon, metal nitrides, and phosphides (like $Ta_3N_5$ and GaP), have been explored for potential use in PEC systems [10,11].

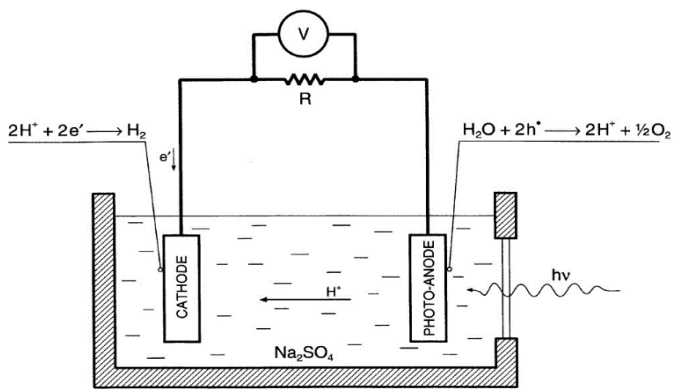

**Fig. 6.** Structure of photo-electrochemical cell (PEC) for water photo-electrolysis [12].

## 2.4 Biomass Gasification

### 2.4.1 Supercritical water gasification

Hydrothermal biomass gasification uses supercritical water to break down organic compounds. The goal is complete conversion to hydrogen, employing glucose as a model biomass compound (Equation 1). Carbon monoxide is the main byproduct, reacting with water to yield carbon dioxide and hydrogen (Equation 2). Abundant water in the system shifts the equilibrium toward hydrogen. Water is crucial not only in Equation 2 but also in biomass hydrolysis to glucose. To expedite the slow water gas shift reaction, alkali metals can be introduced as catalysts [13]. Hydrogen formation is favored at temperatures significantly larger than the critical point of water (Tc = 374 °C and Pc = 22.1 MPa) because of its endothermic nature [13]. Reactions (1) and (2) consume water and are favored at high dilution, allowing hydrogen formation at low temperatures even with very low feedstock concentrations. This principle is utilized in Aqueous Phase Reforming with catalysts like nickel. However, solid catalysts hinder solid biomass utilization due to slow hydrolysis. Significant hydrogen production is possible when using dissolved organic compounds like glucose, sorbitol, or glycol as feedstock [13].

$$C_6H_{12}O_6 + 6H_2O \rightarrow 6CO_2 + 12H_2 \quad (1)$$

$$CO + H_2O \rightleftharpoons CO_2 + H_2 \quad (2)$$

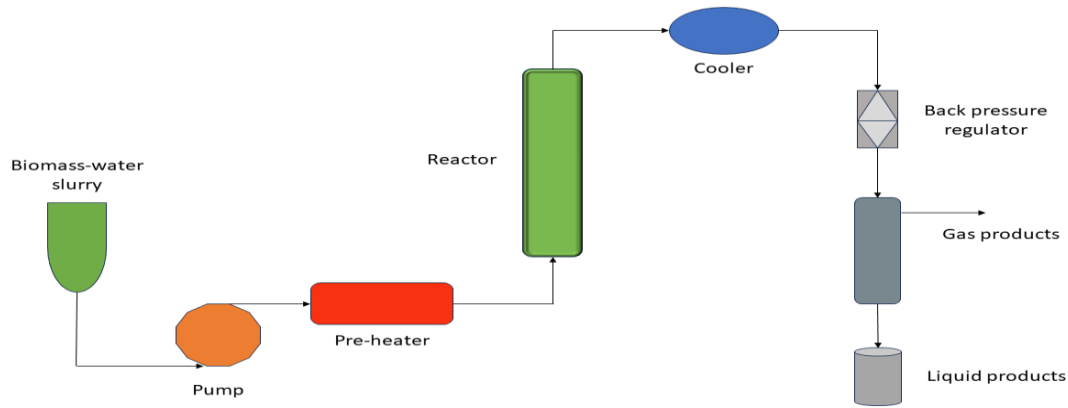

**Fig. 7.** Schematic of Supercritical Water Gasification.

### 2.4.2 Steam gasification

Steam gasification is considered a viable thermochemical approach for hydrogen production, as it is believed to be an efficient means of generating clean, renewable hydrogen while minimizing carbon emissions in the environment [14]. Biomass steam gasification transforms carbon-rich materials into permanent gases (such as $H_2$, CO, $CO_2$, $CH_4$, and light hydrocarbons), as well as producing char and tar. This process can be described by a simplified reaction equation.

Biomass + Steam $\rightarrow$ $H_2$ + CO + $CO_2$ + $CH_4$ + Light and heavy HC + Tar + Char

Hydrogen yield in steam gasification surpasses that in fast pyrolysis followed by char steam reforming. Comparatively, there is no established comparison between steam gasification and supercritical water gasification in terms of hydrogen yield. Steam gasification generates fewer amounts of char and tar due to increased water gas reactions. This technology is ideal for biomass with moisture content below 35% [14]. In general, most biomass sources contain approximately 35% moisture, making steam gasification an excellent choice for biomass processing. This technology is both well-proven and firmly established. Steam gasification boasts several advantages, including its effectiveness in renewable hydrogen production, its ability to yield the highest amount of hydrogen from biomass, and its capacity to generate cleaner products with minimal environmental consequences [14].

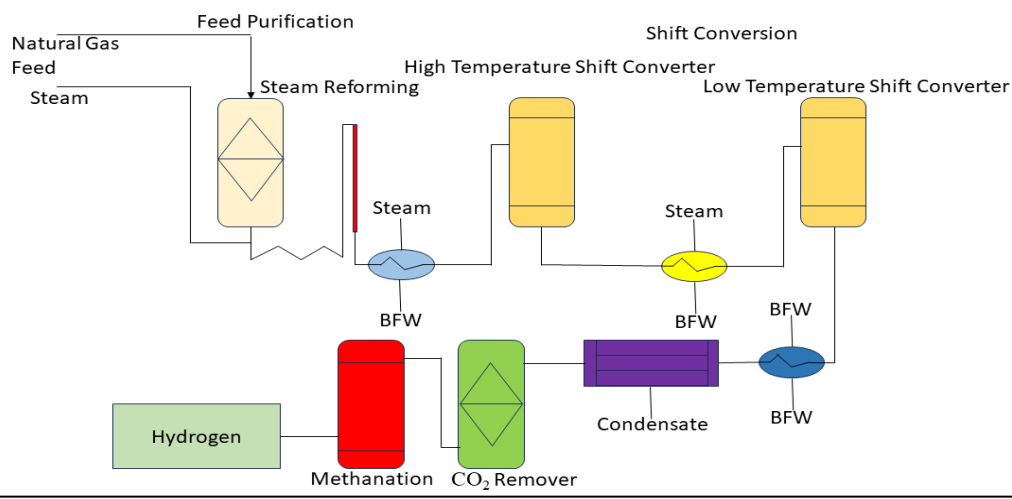

**Fig. 7.** Schematic of Steam Gasification.

## 3. Future Aspects of Hydrogen

The notion of a 'hydrogen economy' was originally conceptualized by John Bockris in the 1970s. It outlined a plan wherein hydrogen is created by electrolyzing water and then transported via pipes to residences, businesses, and gasoline stations, where it is regenerated into power using on-site fuel cells [15]. The current hydrogen economy will expand to meet future demand; the hydrogen economy is not some far-off idea. With the use of heavy-duty, long-range fuel cell cars, hydrogen has the potential to contribute to the decarbonization of the transportation, building, and heating sectors. It can also be blended with natural gas pipelines to provide heating. If hydrogen is utilized to make electro-fuels, or e-fuels, for aviation applications and to provide high-grade heat for industrial processes, its position in the transportation sector may grow over time. By now, it will also permeate the power industry, acting as a seasonal energy storage mechanism and lessening the reduction of renewable energy to enable almost complete decarbonization of the energy sector. According to the International Energy Agency (IEA), there was about 90 Mt $H_2$ demand worldwide [3].

**3.1 Hydrogen as a Chemical Feedstock**

Although the hydrogen economy is currently in position, it is not completely sustainable. The production of hydrogen feedstocks for chemical synthesis and other industrial processes accounts for around 17% (1.1 Gt) of the 6.3 Gt of worldwide energy-related carbon emissions that are caused by the industrial sector annually [15]. Although electrolysis is the only completely green pathway to meet the current hydrogen demand of 90 Mt in this sector for chemical synthesis involving hydrogen feedstocks, since electrification cannot directly impact these processes, "blue" hydrogen—which combines gray hydrogen with carbon capture and storage (CCS) technologies—is a low-carbon alternative [16]. Hydrogen feedstocks are used in widely recognized and expanding processes like methanol generation, ammonia synthesis, and oil refining. In the industry sector, demand is a little bit greater (over 50 Mt $H_2$), mostly for feedstock [3]. About 45 Mt $H_2$ of the need is satisfied by chemical manufacturing, of which about 75 percent is used to produce ammonia and 25 percent is used to produce methanol [3]. In the process of manufacturing steel, direct reduced iron (DRI) uses up the remaining 5 Mt $H_2$ [3]. Industry is the largest end-use sector, accounting for 38% of total final energy demand and 26% of $CO_2$ emissions from the global energy system [3]. Hydrogen is produced using 6% of the industry's total energy consumption. It is mostly employed as a reducing agent in the production of iron and steel and as a feedstock for chemical production [3]. The yearly demand for hydrogen in the industry is 51 Mt [3].

**3.2 Transition Fuel for Transportation**

Over 20% of the world's greenhouse gas emissions and 25% of total energy consumption are attributable to the transportation industry, which gets 90% of its energy from oil products [3]. The sector has used very little hydrogen up until 2020; it made up less than 0.01% of total energy used [3]. Nevertheless, there may be ways to lower emissions with hydrogen and hydrogen-based fuels. Fuel cell electric vehicles (FCEVs) are already offered for sale as passenger cars. FCEVs only emit water vapor and utilize hydrogen and oxygen from the surrounding air [16]. The number of hydrogen refueling stations (HRSs) is increasing more slowly than that of fuel cell electric vehicles (FCEVs), with an average annual growth rate of over 20% between 2017 and 2020. Consequently, the ratio of FCEVs to HRSs is rising, especially in the nations with the largest FCEV sales [3]. In contrast to Japan, where it was just 30:1, this ratio was around 200:1 in 2020 in Korea and 150:1 in the US [3]. This signifies, in part, redundant HRS capacity, as stations are being constructed with the expectation of FCEV growth.

## 4. Conclusion

In conclusion, the text underscores the urgency of transitioning to sustainable energy sources. Current fossil fuel-dependent energy consumption is unsustainable from environmental, economic, and social standpoints. The text introduces the concept of a "hydrogen economy" as a promising solution. Hydrogen, with its high energy density and versatility, is recognized as a key alternative energy carrier for a sustainable future. Various hydrogen production methods, including electrolysis, thermochemical processes, and biomass gasification, are explored and studies shows that electrolysis is the best technique to produce hydrogen due to its high efficiency of --% and cost effectiveness, and we can make it sustainable by using solar energy, photonic energy or methane as energy source. It highlights the significant potential of hydrogen in sectors such as transportation and power generation. However, the transition to a hydrogen-based economy requires greener production methods and addressing environmental concerns. In summary, the text advocates for a sustainable energy shift with a focus on hydrogen-based technologies and the imperative for research, development, and investment in green hydrogen production.